\documentclass[dvips,a4paper,12pt]{article}
\usepackage{graphicx}
\usepackage{amsmath}
\usepackage{authblk}
\usepackage[numbers,sort&compress]{natbib}
\usepackage[bookmarks=true,colorlinks=true,linkcolor=black,citecolor=blue,breaklinks]{hyperref}
\usepackage{bm}
\usepackage{gensymb} 
\newcommand{\be}{\begin{equation}}
\newcommand{\en}{\end{equation}}
\newcommand{\bea}{\begin{eqnarray}}
\newcommand{\ena}{\end{eqnarray}}
\newcommand{\lbl}[1]{\label{eq:#1}}
\newcommand{\lbltab}[1]{\label{tab:#1}}

\newcommand{\rf}[1]{(\ref{eq:#1})}
\newcommand{\Table}[1]{\ref{tab:#1}}
\newcommand{\fig}[1]{\ref{fig:#1}}

\newcommand{\braque}[1]{{\langle #1 \rangle}}
\newcommand{\bc}{\begin{center}}
\newcommand{\ec}{\end{center}}
\newcommand{\bt}{\begin{tabular}}
\newcommand{\et}{\end{tabular}}
\newcommand{\ba}{\begin{array}}
\newcommand{\ea}{\end{array}}

\newcommand{\fpid}{F_\pi^2}
\newcommand{\fpi}{F_\pi}
\newcommand{\piz}{{\pi^0}}
\newcommand{\pip}{{\pi^+}}
\newcommand{\pim}{{\pi^-}}
\newcommand{\Kp}{{K^+}}
\newcommand{\Km}{{K^-}}
\newcommand{\Kz}{{K^0}}
\newcommand{\Kzb}{{\bar{K}^0}}
\newcommand{\mpi}{m_\pi}
\newcommand{\mpid}{m_\pi^2}
\newcommand{\im}{{\rm Im\,}}

\newcommand{\mpiz}{m_\pi}
\newcommand{\mpizd}{m_\pi^2}
\newcommand{\alphatilde}{\tilde{\alpha}}
\newcommand{\lp}{{\ell^+}}
\newcommand{\lm}{{\ell^-}}
\newcommand\T{\rule{0pt}{3.0ex}}         
\newcommand{\arctanh}{\hbox{arctanh}}
\newcommand{\gapprox}{%
\mathrel{%
\setbox0=\hbox{$>$}\raise0.23ex\copy0\kern-\wd0\lower0.60ex\hbox{$\sim$}}}

\begin{document}

\begin{flushright}
{LAL-15-382}
\end{flushright}

\vspace{1cm}
\begin{center}
{\LARGE\bf Single  pion contribution to the hyperfine
  splitting in muonic hydrogen}
\vspace{1cm}

{\large Nguyen Thu Huong}$^{\hbox{a}}$ ,
{\large Emi Kou}$^{\hbox{b}}$ 
{\large and Bachir Moussallam}$^{\hbox{c}}$ 
\vspace{0.5cm}

$^{\hbox{a}}$ {\small Faculty of Physics, VNU University of Science,
    Vietnam National University, 334 Nguyen Trai, Thanh Xuan, Hanoi,
    Vietnam}\\ 
$^{\hbox{b}}${\small Laboratoire de l'Acc\'el\'erateur Lin\'eaire,
  Univ. Paris-Sud, CNRS/IN2P3, Universit\'e Paris-Saclay, 91898 Orsay
  C\'{e}dex, France} \\
$^{\hbox{c}}${\small Groupe de physique th\'eorique, IPN, Universit\'e
    Paris-Sud 11, 91406 Orsay, France}

\vspace{0.5cm}

\today

\vspace{0.5cm}

\begin{minipage}{0.90\linewidth}
\centerline{\bf Abstract}

{\small
\indent A detailed discussion of the long-range one-pion exchange (Yukawa
potential) contribution to the 2S hyperfine splitting in muonic hydrogen which
had, until recently, been disregarded is presented. We evaluate the relevant
vertex amplitudes, in particular $\pi^0\mu^+\mu^-$, combining low energy
chiral expansions together with experimental data on $\pi^0$ and $\eta$ decays
into two leptons. A value of $\Delta{E}^\pi_{HFS}= -(0.09 \pm
0.06)\ \mu\rm{eV}$ is obtained for this contribution.}
\end{minipage}

\end{center}

\section{Motivation}
The first accurate measurement of the $2S_{1/2}^{F=1}-2P_{1/2}^{F=2}$ Lamb shift
transition in muonic hydrogen~\cite{Pohl:2010zza} has led, with the help of
the currently accepted theoretical formulae
(e.g.~\cite{Pachucki:1996zza,Borie:2004fv}), to a determination of the proton
radius $r_E$ with a precision of 0.8 per mil. The proton size puzzle arose
from the discrepancy, by five standard deviations, between this result and the
{CODATA-2010} value{~\cite{Mohr:2012tt}}, which was based on
ordinary hydrogen spectroscopy as well as $ep$ scattering.
This has stimulated a number of new theoretical and experimental
investigations (see e.g. the review~\cite{Carlson:2015jba}). In particular,
Antognini at al.~\cite{Antognini:1900ns} have measured both the
{$\nu_t\equiv 2S_{1/2}^{F=1}-2P_{3/2}^{F=2}$ and the $\nu_s\equiv
  2S_{1/2}^{F=0}- 2P_{3/2}^{F=1}$} transitions which has confirmed and refined
the previous result on the Lamb shift (increasing the $r_E$ discrepancy to
$7\sigma$) and further provides an experimental value for the $2S$ hyperfine
splitting\footnote{{The 2S
hyperfine splitting is extracted from the experimental measurements through
equation, $\Delta E^{\rm 2S}_{\rm HFS}= h\nu_s-h\nu_t+
\Delta E_{\rm HFS}^{2P_{3/2}}-\delta$ where $h$ is the Planck constant and the 2P
hyperfine splitting $\Delta E_{\rm HFS}^{2P_{3/2}}$ and the 2P $F=1$ mixing
parameter $\delta$ are computed theoretically~\cite{Borie:2004fv,
   Martynenko:2006gz}.}} 
\be\lbl{EHFSexp}
\Delta E_{HFS}^{exp}=22.8089(51)\ (\hbox{meV})\ .
\en
The hyperfine splitting is interesting as it probes aspects of the proton
structure somewhat differently from the Lamb shift.  While the influence of
the proton radius $r_E$ is suppressed, the main structure dependent
contribution is proportional to the Zemach radius $r_Z$: $\Delta
E_{HFS}^Z=-0.1621(10)\,r_Z$ meV (with $r_Z$ in fm), as given in the 
review~\cite{Antognini:2013jkc}, and the next main structure
dependent contribution is that associated with the forward proton
polarizabilities. It has been estimated in ref.~\cite{Carlson:2008ke} as:
$\Delta E_{HFS}^{pol}= (8.0\pm2.6)$ $\mu\hbox{eV}$ (see also
~\cite{Faustov:2002yp}). It is noteworthy that the value of $r_Z$ that one
determines from the HFS measurement in muonic hydrogen: $r_Z=1.082(37)$
fm~\cite{Antognini:1900ns} is in agreement with the value computed in terms of
the proton form factors $G_E$, $G_M$ measured in $ep$ scattering, 
$r_Z=1.086(12)$ fm~\cite{Nazaryan:2005zc}, at the present level of accuracy.

A possible role in muonic hydrogen of light, exotic (universality violating)
particles, with vector or axial-vector ($J^{PC}=1^{--},\, =1^{++}$) quantum
numbers has been
considered~\cite{Barger:2010aj,Karshenboim:2014tka}. Similarly, the influence
of exchanging a light pseudo-scalar particle ($J^{PC}=0^{-+}$) was recently
studied in ref.~\cite{Keung:2015qla}. In that case, the HFS splitting is
affected but not the (appropriately defined) Lamb shift. 

\begin{figure}
\centering
\includegraphics[width=0.50\linewidth]{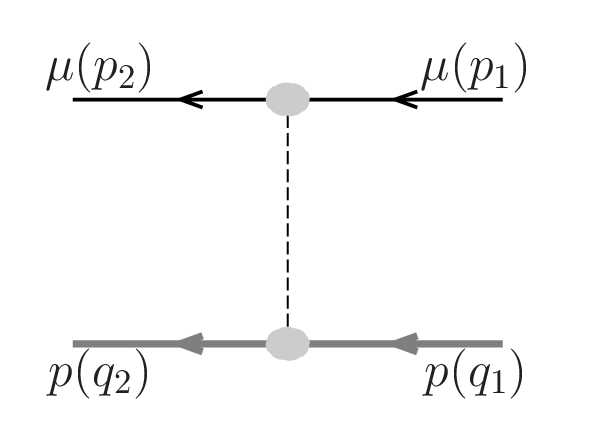}
\caption{\small Single pion exchange in the $\mu p\to \mu p$ amplitude}
\label{fig:Yukawa}
\end{figure}

In this note, we point out that a light pseudo-scalar particle exists within
the standard model, the neutral pion, and we perform the exercise to estimate
the influence of the one-pion exchange mechanism on $\Delta{E}_{HFS}$. We will
show that using chiral symmetry allows one to evaluate the two vertex
functions which are needed, represented by blobs in Fig.~\fig{Yukawa}, for
small momentum transfer, based on experimental data.

The coupling of the $\pi^0$ to a lepton pair proceeds (within the standard
model) via two virtual photons. The $\mu p\to \mu p$ one-pion exchange
amplitude can also be viewed as a two-photon exchange amplitude. The pion pole
in the Compton amplitude $\gamma p\to \gamma p$ contributes to the so-called
proton backward spin polarizability $\gamma_\pi$
(e.g.~\cite{Drechsel:2007sq}). The corresponding contribution in muonic
hydrogen is then expected to be suppressed by one power of $\alpha$ as
compared to the forward proton polarizability contribution. This explains why
the simple mechanism of Fig.~\fig{Yukawa} does not seem to have been
previously considered until very
recently~\cite{Zhou:2015bea,Hagelstein:2015lph}. 
Some enhancement  might be expected from the
fact that $\gamma_\pi$ is numerically large compared to the forward
polarizabilities $\alpha_p$, $\beta_p$ and from the fact that the 
Yukawa potential has a relatively long range (on the scale of the proton size)
which increases the overlap with the atomic wave-functions
As a final motivation, let us recall that the $\pi^0 \mu^+ \mu^-$ coupling
plays a significant role among the hadronic contributions to the muon
$g-2$~\cite{Knecht:2001qf} and it is thus of interest to probe the level of
sensitivity of muonic hydrogen to this coupling.

\section{Pion coupling amplitudes to leptons and to nucleons}
\subsection{$\pi^0$-lepton coupling}
\begin{figure}
\centering
\includegraphics[width=0.40\linewidth]{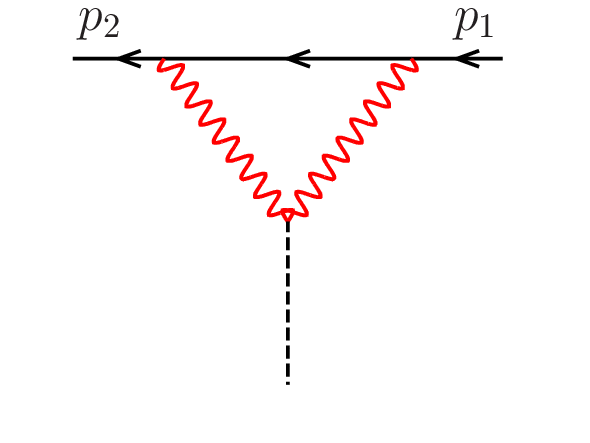}\includegraphics[width=0.40\linewidth]{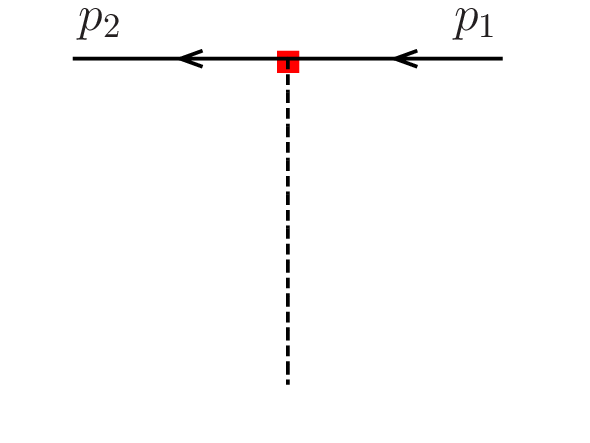}
\caption{\small Feynman graphs which generate the $\pi^0$-lepton vertex
  amplitude at  leading order in the chiral expansion.}
\label{fig:pillvertex}
\end{figure}
For low momentum transfer, the $P \lp \lm$ vertex amplitude, where $P$
is a light neutral pseudo-scalar meson ($\pi^0$ or $\eta$) and
$\ell^\pm$ is a light lepton ($e^\pm$ or $\mu^\pm$), can be evaluated
in the chiral expansion\footnote{We consider here the coupling
  mediated by the electromagnetic interaction. The coupling mediated
  by the weak interaction is comparatively suppressed by two orders of
magnitude.}~\cite{Savage:1992ac}. At leading order, the
amplitude is given from the two diagrams shown on
fig.~\fig{pillvertex}. In the one-loop diagram, the $P\gamma\gamma$
vertex is generated by the Wess-Zumino-Witten Lagrangian
(see~\cite{Weinberg:1996kr}, chap.~22)
\be
{\cal L}_{WZ}= \frac{\alpha}{8\pi\fpi}\epsilon^{\mu\nu\alpha\beta} \big(
\pi^0 + {1\over\sqrt3} \eta  \big)\,F_{\mu\nu}F_{\alpha\beta}
\en
with the sign corresponding to the convention $\epsilon^{0123}=1$ (we
also use $\gamma^5=i\gamma^0\gamma^1\gamma^2\gamma^3$). This diagram
accounts for the contributions of photons with low energy compared to
1 GeV. The higher energy contributions are parametrized through two
chiral coupling constants $\chi_1$, $\chi_2$ in the
Lagrangian~\cite{Savage:1992ac},
\be
{\cal L}_{SLW}=\frac{3i{\alpha^2}}{32\pi^2}\,
\bar{\ell}\gamma^\mu\gamma^5 \ell\,\left(
\chi_1\,\braque{
 Q^2 U^\dagger D_\mu U
-Q^2 U D_\mu U^\dagger}
+\chi_2\,\braque{
 Q U^\dagger Q D_\mu U
-Q U Q D_\mu U^\dagger }
\right)
\en
where $U$ is the chiral $SU(3)$ matrix,
\be
U=\exp\frac{i\Phi}{\fpi},\quad
\Phi=\begin{pmatrix}
\pi^0+\dfrac{\eta}{\sqrt3}& \sqrt2\,\pip &\sqrt2\,\Kp\\
\sqrt2\,\pim & -\pi^0+\dfrac{\eta}{\sqrt3} & \sqrt2\,\Kz\\
\sqrt2\,\Km & \sqrt2\,\Kzb & -\dfrac{2\,\eta}{\sqrt3}\\
\end{pmatrix}
\en
and
\be\lbl{DmuU}
D_\mu U=\partial_\mu U-i(v_\mu+a_\mu)\,U +iU\,(v_\mu-a_\mu)
\en
where $v_\mu (\,a_\mu$) are external vector (axial-vector) sources
(see~\cite{gl85}) and $Q$ is the charge matrix,
$Q=diag(2/3,-1/3,-1/3)$. The tree graph shown in fig.~\fig{pillvertex}
is computed from this Lagrangian. The coupling constants $\chi_1$,
$\chi_2$ remove the ultraviolet divergence of the one-loop graph.
Assuming the leptons to be on their mass shell,
the $P \lp \lm$ vertex amplitude can  be expressed in terms of a single
Dirac structure, 
\be\lbl{pillvertex}
i{\cal T}_{P \lp\lm}= r_P\frac{\alpha^2 m_\ell}{2\pi^2\fpi}\, 
{\cal A}_\ell( (p_1-p_2)^2 )\,\bar{u}_\ell(p_2)\gamma^5 u_\ell(p_1) ,
\en
where $r_P=1,\ 1/\sqrt{3}$ if $P=\pi,\, \eta$. 
In practice, dimensional regularization brings in some scheme dependence 
because of the presence of the $\gamma^5$ matrix. For instance, the amplitudes
computed in refs.~\cite{Savage:1992ac} and~\cite{Knecht:1999gb} differ by a
constant. Some discussion of this point can be found in
ref.~\cite{RamseyMusolf:2002cy}. For definiteness, we will choose the
convention of~\cite{Knecht:1999gb}, which gives ${\cal A}_\ell(s)$ in the form
\be\lbl{A(s)}
{\cal A}_\ell(s)=\chi_P(\Lambda)+{3\over2}\log\frac{m_\ell^2}{\Lambda^2} 
-\frac{5}{2} +C_\ell(s),\quad
\chi_P=-{1\over4}(\chi_1+\chi_2)\ 
\en
with
\be\lbl{C_l}
C_\ell(s)=\frac{1}{\beta_\ell(s)}\left[
\hbox{Li}_2\frac{\beta_\ell(s)-1}{\beta_\ell(s)+1}
+\frac{\pi^2}{3}
+\frac{1}{4}\log^2 \frac{\beta_\ell(s)+1}{\beta_\ell(s)-1}
\right],\quad
\beta_\ell(s)=\sqrt{1-4m_\ell^2/s}\ .
\en
Using $\overline{MS}$ renormalization, the coupling constant
combination $\chi_P$ becomes scale
dependent with $ d/d\Lambda\chi_P(\Lambda)=3/\Lambda$, which ensures that
${\cal A}_\ell$ is scale independent. 

The value of $\chi_P(\Lambda)$ must be
determined from experiment. For this purpose, we can use either $\pi^0\to e^+
e^-$ which was measured recently by the KTeV
collaboration~\cite{Abouzaid:2006kk} or $\eta \to \mu^+ \mu^-$
(see~\cite{Agashe:2014kda}). It is convenient to consider the ratio
$R_P=\Gamma(P\to \lp \lm)/\Gamma(P\to \gamma\gamma)$ which should be less
sensitive to higher order chiral corrections than the individual modes. It is
expressed as follows, in  terms of the amplitude ${\cal A}_\ell$,
\be
R_P= 
\frac{2\,\alpha^2 m_\ell^2}{\pi^2 m_P^2}
 \,\beta_\ell(m^2_P)\,\left\vert
{\cal A}_\ell(m^2_P)
\right\vert^2\ \ .
\en
In the case of the $\pi^0$,
the quantity measured experimentally is the branching ratio for the
decay mode $\pi^0 \to e^+ e^- (\gamma)$, including photons in the
final state such that $s_{e^+e^-} \ge 0.95 m^2_\piz$.  The ratio which
interest us, $R_\piz$, can be deduced from this result by removing the
bremsstrahlung and the associated radiative corrections. These have been
revised recently in ref.~\cite{Vasko:2011pi}. Using the results
of that work, one deduces
\be
R_\piz^{exp}=(6.96\pm0.36)\cdot10^{-8}\ .
\en
There are two solutions for $\chi_P$ which correspond to this experimental
result 
\be\lbl{chisolspi0}
\ba{l}
a)\, \chi_P(m_\rho)= 4.51\pm 0.97\qquad \\[2pt]
b)\, \chi_P(m_\rho)=-19.41\pm 0.97\\
\ea\en
(in which the scale  was set to $\Lambda=m_\rho=0.774$ GeV). In order to
decide on which solution to choose, we can compare with the model proposed in
ref.~\cite{Knecht:1999gb}. It is based on a rigorous sum rule which holds in
the large $N_c$ limit of QCD and the approximation of retaining only the
lightest resonance in the sum. This model gives,
\be\lbl{ChiLMD}
\chi_P^{LMD}(\Lambda)=\frac{11}{4}
                     -\frac{3}{2}\log{m_\rho^2\over\Lambda^2}
                     -\frac{4\pi^2\fpid}{m_\rho^2}
\en
and the uncertainty was estimated in ref.~\cite{Knecht:1999gb} to be of the
order of 40\%. Thus, one has 
\be
\chi_P^{LMD}(m_\rho)\simeq 2.2\pm 0.8 \ .
\en 
This result lies within one sigma of solution {$a)$} and is not compatible
with solution {$b)$}. This argument  suggests that solution {$a)$}
is more likely to be the physically correct one. 

Alternatively, we can 
determine the coupling constant {$\chi_P$} from the decay mode of the
$\eta$ meson, $\eta\to \mu^+\mu^-$ for which the experimental
branching fraction is (see~\cite{Agashe:2014kda}):
$BF(\eta\to\mu^+\mu^-)= (5.8\pm0.8)\cdot10^{-6}$ leading to
\be
R_\eta^{exp}= (1.47\pm0.20)\cdot10^{-5}
\en
There are again two solutions for $\chi_P$ corresponding to this experimental
result, 
\be\lbl{chisolseta}
\ba{l}
a')\, \chi_P(m_\rho)= 1.69\pm 0.87\qquad\\[2pt]
b')\, \chi_P(m_\rho)= 7.96\pm 0.87\ .
\ea\en
None of these solutions is compatible with $b)$ of eq.~\rf{chisolspi0}: one can
therefore safely conclude that  solution $b)$ must be eliminated. We can
also eliminate $b^{\prime})$ which is not compatible with the model
estimate~\rf{ChiLMD} while $a^{\prime})$ is. It seems reasonable, for our
purposes, to perform  an average of the {$a)$} and {$a^{\prime})$} values and
thus use
\be\lbl{Chipaverage}
\chi_P(m_\rho)= 3.10\pm 1.50\ ,
\en
where we have slightly rescaled the error such that the two central values of
$a)$ and $a^{\prime})$ lie within the error.

\subsection{$\pi^0$-proton coupling}
At leading order in the chiral expansion, the pion-nucleon coupling is
given, at tree level, from the chiral Lagrangian~\cite{Gasser:1987rb}
\be\lbl{piNNlag}
{\cal L}_{\pi NN}=\bar{\psi}\left(i\gamma^\mu  \Delta_\mu -m_N +i{g_A\over2}
\gamma^\mu\gamma^5 u^\dagger D_\mu U u^\dagger\right) \psi,\quad
\en
where $U$  is the $SU(2)$ chiral matrix here, $u=\sqrt{U}$, and
\be
\Delta_\mu= \partial_\mu +\Gamma_\mu,\quad
\Gamma_\mu= \frac{1}{2}[u^\dagger,\partial_\mu u]
-\frac{1}{2}i u^\dagger(v_\mu+a_\mu) u
-\frac{1}{2}i u (v_\mu-a_\mu)u^\dagger\ 
\en
$v_\mu$ ($a_\mu$) being external vector (axial-vector) sources and $\psi$ is an
isospin spinor containing the proton and the neutron,
\be
\psi=\begin{pmatrix}
\psi_p\\
\psi_n\\
\end{pmatrix}\ .
\en
The coupling constant $g_A$ in the Lagrangian.~\rf{piNNlag} is easily
identified as the axial charge of the proton and also controls the
neutron-proton matrix element of the charged axial current,
\be\lbl{gAmat}
\lim_{q'=q}\braque{p(q')\vert \bar{u}\gamma^\mu\gamma^5d \vert n(q)}
=g_A\,\bar{u}_p(q) \gamma^\mu \gamma^5 u_n(q)\ .
\en
It is determined  from  neutron beta decay experiments to have the
following positive\footnote{The absolute value of $g_A$ is obtained from the
neutron lifetime and its sign, we remind, is unambiguously
determined from the asymmetry parameter of the neutron beta decay
which, using eq.~\rf{gAmat}, is given by: $A=2(g_A-g_A^2)/(1+3g_A^2)$. The
experimental value is~\cite{Agashe:2014kda} $A=-0.1184(10)$.}
value~\cite{Agashe:2014kda}
\be
g_A=1.2723\pm0.0023
\en
The pion-proton vertex amplitude is then deduced from the
Lagrangian~\rf{piNNlag} to be
\be\lbl{piNNvertex}
i{\cal T}_{\pi pp}= -g_{\pi pp} \,\bar{u}_p(q_2)\gamma^5 u_p(q_1) ,\qquad
g_{\pi pp}= \frac{g_A m_p}{\fpi}
\en
The expression of the coupling constant $g_{\pi pp}$  at leading chiral
order, in terms of $g_A$, $m_p$ and $\fpi$, as it appears in the above
expression is, of course, the content of the Nambu-Goldberger-Treiman
relation (e.g.~\cite{Weinberg:1996kr} chap.~19). It is known that the
higher order chiral corrections to this relation do not exceed a few
percent.

\section{Energy shifts in muonic hydrogen }
\subsection{ $q^2=0$ approximation}
Having determined the $\pi^0\mu\mu$ vertex (eq.~\rf{pillvertex}) and the
$\pi^0 pp$ vertex (eq.~\rf{piNNvertex}) it is straightforward to derive 
the muon-proton scattering amplitude, $\mu(p_1)
p(q_1)\to \mu(p_2) p(q_2)$ associated with one-pion exchange
(Fig.~\fig{Yukawa}) 
\be
{\cal T}_{\mu p}= -{4m_\mu m_p\, \alpha^2 g_A {\cal A}_\mu((p_1-p_2)^2)
\over 8\pi^2 \fpid }
{\bar{u}_\mu(p_2)\gamma^5 u_\mu(p_1)\, 
\bar{u}_p(q_2)\gamma^5 u_p(q_1)\over
(p_1-p_2)^2 -\mpizd}
\en 
For our purposes, we can consider that both the muon and the proton
are non-relativistic, therefore 
\be
(p_1-p_2)^2=(q_1-q_2)^2\simeq
-(\vec{p}_1-\vec{p}_2)^2\equiv-q^2 \ . 
\en
At first, let us make the approximation to set $q^2=0$ in the vertex function
${\cal A}_\mu$. We then obtain the non-relativistic Yukawa potential in
momentum space, 
\be\lbl{yukawamom}
V_{\mu p}(\vec{q})= -\frac{{\cal T}_{\mu p}}{4 m_\mu m_p}=\lambda\,
{\cal A}_\mu(0)\,
\frac{\vec{\sigma}_\mu\cdot\vec{q} \,\, \vec{\sigma}_p\cdot\vec{q}}
{q^2 + \mpizd},\quad 
\lambda= {\alpha^2  g_A\over 8\pi^2 \fpid}
\en
The contributions to the atomic energy shifts are most easily
performed by Fourier transforming to configuration space,
\be\lbl{Vmup}
V_{\mu p}(\vec{r})=\tilde{\lambda}\, \left[
\vec{\sigma}_\mu\cdot \vec{\sigma}_p\, V_{SS}(\vec{r})
+ S_{12}\, V_T(r)
\right] , \quad
\tilde{\lambda}=-\lambda\,{\cal A}_\mu(0)\,\frac{\mpizd}{12\pi}
\en
where $S_{12}=3\,\vec{\sigma}_\mu\cdot
\hat{r}\, \vec{\sigma}_p\cdot \hat{r}
-\vec{\sigma}_\mu\cdot\vec{\sigma}_p$  is the so-called tensor
operator, and
\bea\lbl{VssVt}
&& V_{SS}(\vec{r})=\frac{\exp(-\mpiz r)}{r}
-\frac{4\pi}{\mpizd}\delta^3(\vec{r}),\quad
\nonumber\\
&& V_T(r)=\left(
1+\frac{3}{\mpiz r} +\frac{3}{\mpizd r^2}
\right)\,\frac{\exp(-\mpiz r)}{r}\ .
\ena
Making use of
the average result~\rf{Chipaverage} for $\chi_P$, one obtains the following
values for ${\cal A}_\mu(0)$ and for the overall coupling $\tilde{\lambda}$ in
muonic hydrogen\footnote{We also use  $\fpi=92.21(14)$
  MeV and $\mpi=m_\piz=134.9766(6)$.}
\be\lbl{lambdatilde}
{\cal A}_\mu(0)=-5.37\pm 1.5\ ,\quad
\tilde{\lambda}= (2.61\pm0.49)\cdot 10^{-7}\ .
\en

We can now compute the energy shifts 
of muonic hydrogen caused by the one-pion exchange amplitude.  
We will consider both the 2S and 2P energy shifts for completeness, 
the relevant radial Coulomb wave-functions are,
\be
\psi_{2S}(r)={1\over\sqrt2}\exp\left(-{\mu\alpha r\over2}\right)
(1 -{\mu\alpha r\over2}),\
\psi_{2P}(r)={1\over2\sqrt6}\exp\left(-{\mu\alpha
  r\over2}\right)\mu\alpha r
\en
where $\mu$ is the muon-proton reduced mass $1/\mu=1/m_\mu+1/m_p$.
From these, one computes the expectation values of the components
$V_{SS}$ and $V_T$ of the Yukawa potential. For the $S$-wave, firstly,
one has
\be
\braque{2S\vert V_{SS}\vert 2S}\equiv Y_S(\mpiz)
=- \frac{ (\mu\alpha)^4}{ \mpiz^3 } \,
\frac{8+11\alphatilde+8\alphatilde^2+2\alphatilde^3}
{4(1+\alphatilde)^4},\quad 
\alphatilde=\frac{\mu\alpha}{\mpiz}\,\ .
\en
When computing the expectation value in the $2S$ state, the
contribution from the delta function in the potential $V_{SS}$ cancels
the leading term in $\alpha$ from the contribution of the first
piece. As a result of this cancellation, $Y_S$ scales as $\alpha^4$
and has a negative sign.  For the 2P states, one has
\bea
&&\braque{2P\vert V_{SS}\vert 2P}\equiv Y_P=\mpiz\alphatilde^5\,\frac{1 }
{4(1+\alphatilde)^4}
\nonumber\\
&&\braque{2P\vert V_T\vert 2P}\equiv T_P=\mpiz\alphatilde^5\,
\frac{5+4\alphatilde+\alphatilde^2}
{8(1+\alphatilde)^4}
\ena
Table~\Table{Eshifts} lists the expressions for the shifts in the 2P
and the 2S states of muonic hydrogen in terms of the integrals $Y_S$, $Y_P$,
$T_P$ and the overall coupling $\tilde{\lambda}$ (given in eqs.~\rf{Vmup} and
~\rf{lambdatilde}) as well as the central numerical values. The contributions
to the $2P_{3/2}$ states is particularly suppressed because the leading terms
in $\alpha$ cancel in the combination $Y_P-{2\over5} T_P$. Finally, in the
$q^2=0$ approximation, the contribution from single pion exchange to the $2S$
hyperfine splitting in muonic hydrogen is
\be\lbl{2SHFSpion}
\Delta E_{HFS}^{\pi} =-(0.19\pm0.05)\ \mu\hbox{eV}\ ,
\en
which is small but not irrelevant. In contrast, the contributions to the HFS
in the 2P states, as can be deduced from table~\Table{Eshifts} are too small
to be of physical relevance. Our result ~\rf{2SHFSpion} disagrees with the one
quoted in ref.~\cite{Zhou:2015bea} which uses the same approximation. We could
trace the origin of the discrepancy, essentially, to an incorrect coefficient
for the delta function in the Yukawa potential.

\begin{table}
\centering
\bt{c|c|c}\hline\hline
State & Expression & Value ($\mu$eV) \\ \hline
\T $2P^{F=2}_{3/2}$ & $\tilde{\lambda}\,\left(Y_P-{2\over5} T_P\right)$ & $-1.3\,10^{-7}$\\
\T $2P^{F=1}_{3/2}$ & $-{5\over3}\tilde{\lambda}\,\left(Y_P-{2\over5}T_P\right)$& $ 2.1\,10^{-7}$\\[2pt]
\multispan 3\dotfill \\
\T $2P^{F=1}_{1/2}$ & $-{1\over3}\tilde{\lambda}\,\left(Y_P-4\,T_P\right)$ & $0.9\,10^{-4}$\\[2pt]
\T $2P^{F=0}_{1/2}$ & $\tilde{\lambda}\,\left(Y_P-4\,T_P\right)$ & $-2.8\,10^{-4}$\\ \hline
\T $2S^{F=1}_{1/2}$ & $\tilde{\lambda}\, Y_S     $ & -0.049\\
\T $2S^{F=0}_{1/2}$ & $-3\,\tilde{\lambda}\, Y_S$ & 0.146\\ \hline\hline
\et
\caption{\small Contributions from the single pion exchange amplitude to the
  $2S$ and the $2P$ energy levels in muonic hydrogen}
\lbltab{Eshifts}
\end{table}

\subsection{Influence of the vertex functions momentum dependence}
\begin{figure}
\centering
\includegraphics[width=0.60\linewidth]{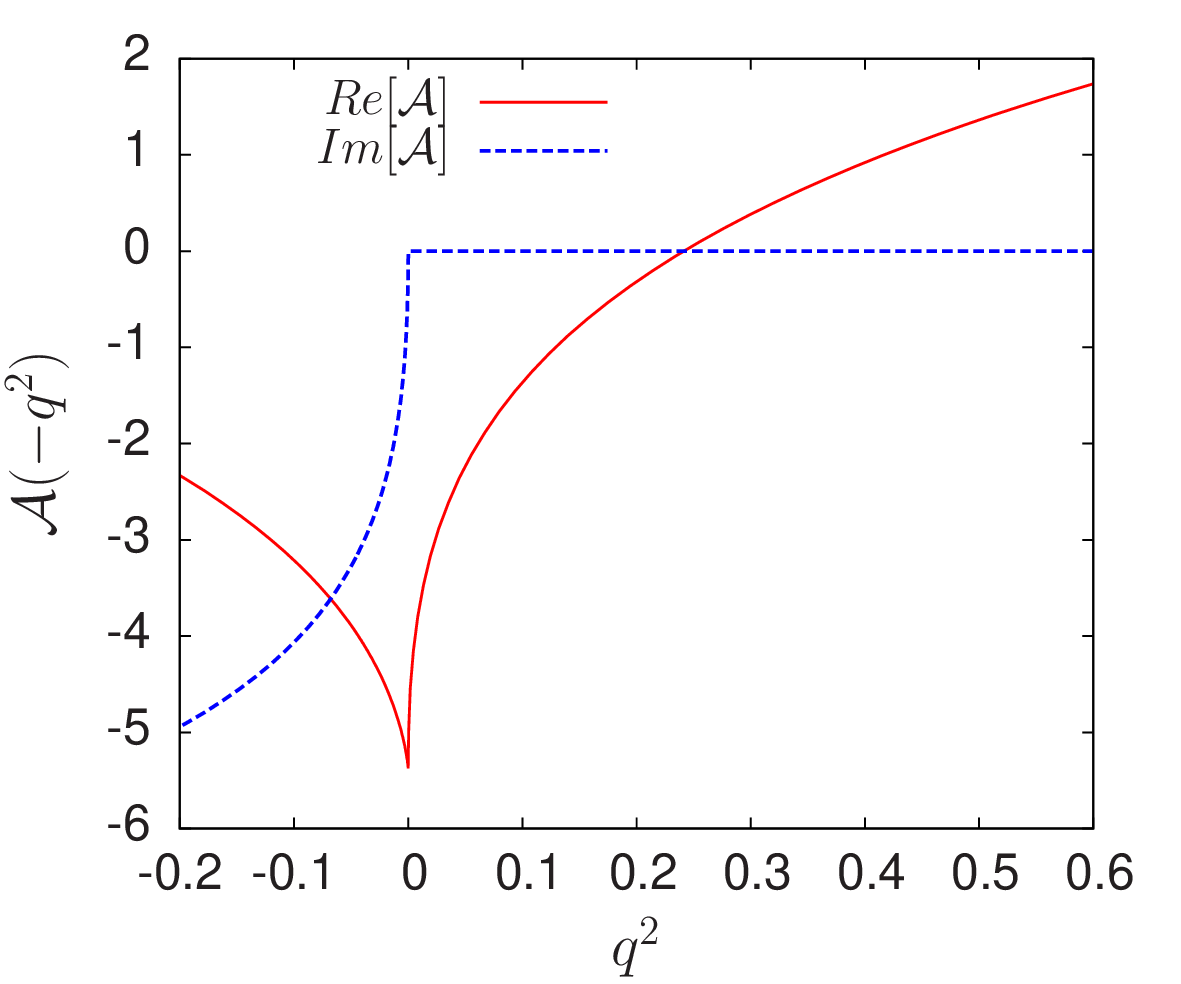}
\caption{\small Vertex function ${\cal A}_\mu$ as a function of $q^2$.}
\label{fig:Aq2}
\end{figure}

\begin{figure}
\centering
\includegraphics[width=0.60\linewidth]{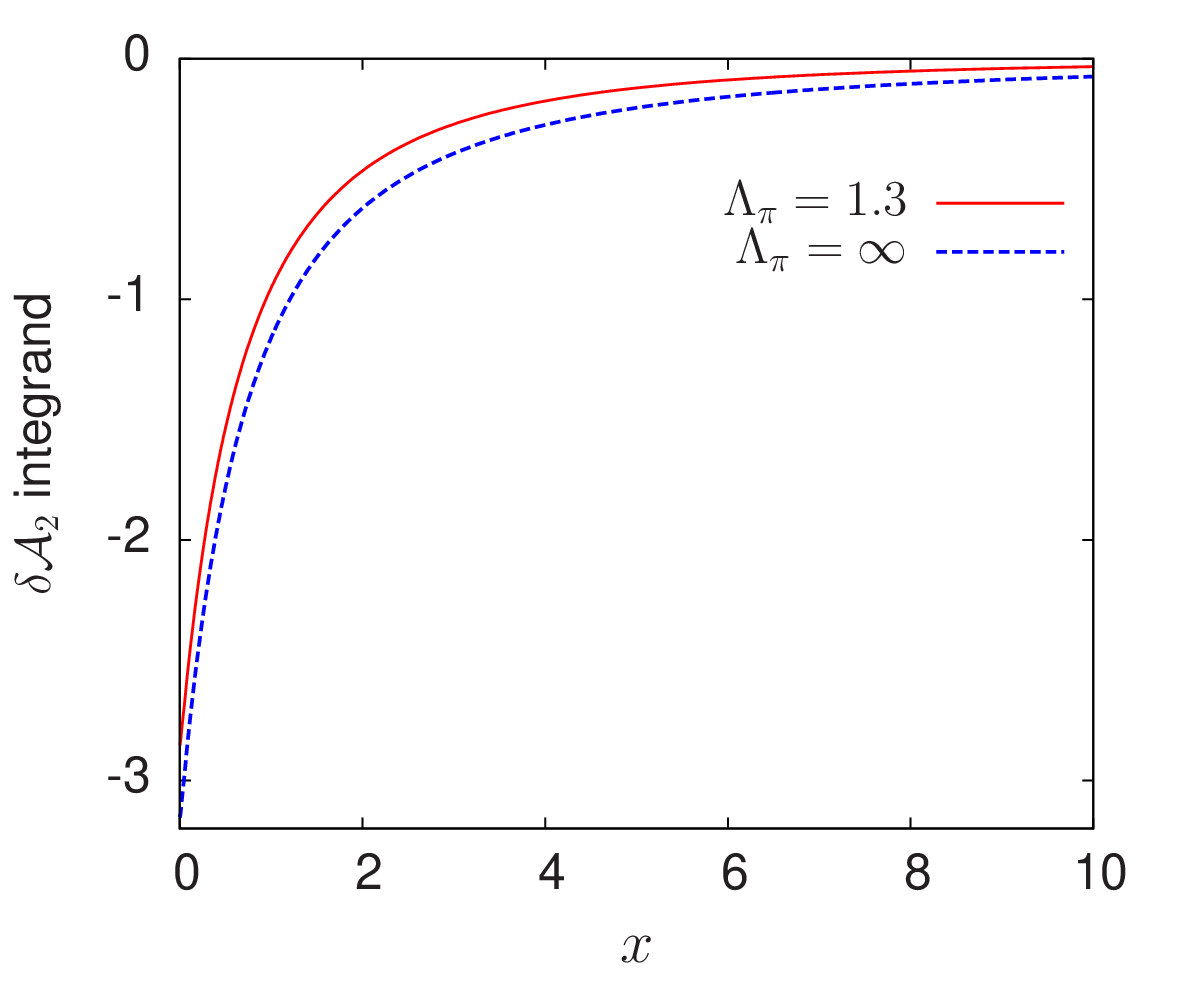}
\caption{\small Integrand of the corrective term $\delta{\cal A}_2$ 
given in  eq.~\rf{deltaA2}.}
\label{fig:FV2}
\end{figure}

The results quoted above were obtained setting $q^2=0$ in the vertex function
${\cal A}_\mu$. It was pointed out in ref.~\cite{Hagelstein:2015lph} that this
is not a good approximation. Plotting ${\cal A}_\mu(-q^2)$ (see
fig.~\fig{Aq2}) shows indeed that the vertex function has a strong cusp at
$q^2=0$ which induces a rapid variation. In the following we evaluate the
corrections induced by the $q^2$ variation of ${\cal A}_\mu$. This is easily
done by using the dispersion relation representation of the function ${\cal
  A}_\mu(-q^2)$,
\be\lbl{disprelA}
{\cal A}_\mu(-q^2)= {\cal A}_\mu(0) -\frac{q^2}{\pi} \int_0 ^\infty ds'\,
\frac{\im {\cal A}_\mu(s')}{s' (s' +q^2) }\ .
\en
For small values of $q^2$ (compared to 1 $\rm{GeV}^2$) we can use the leading
order chiral approximation which gives, for the imaginary
part~\cite{Savage:1992ac}, 
\be
\ba{lll}
\im {\cal A}_l(s')= & -\pi\,\dfrac{\arctan{\sqrt{ 4m_l^2/s'-1}}}
{\sqrt{4m_l^2/s'-1}} &\quad (s' \le 4m_l^2)\\[0.5cm]
\im {\cal A}_l(s')= & -\pi \,\dfrac{\arctanh\sqrt{1-4m_l^2/s'}}
{\sqrt{1- 4m_l^2/s'}}
 &
\quad (s' \ge 4m_l^2)\\
\ea\en
(which is easily verified to be reproduced by the explicit
expression~\rf{A(s)}~\rf{C_l} of ${\cal A}_l$).  Beyond the low $q^2$ region,
estimates of the behaviour of ${\cal A}_\mu$ may be obtained based on
modellings of the
$\pi^0\gamma^*\gamma^*$ form factor
(e.g.~\cite{Dorokhov:2009xs,Masjuan:2015lca} for recent work, see
also~\cite{Ametller:1993we} were a list of references to earlier work can be
found). We will not consider these in detail here and content ourselves with a
simple estimate of the role of the $q^2\gapprox 1$ $\rm{GeV}^2$ region, taking
into account the $q^2$ dependence attached to the $\pi pp$ vertex. In this
case, a weak cusp is expected from the three pions threshold at $q^2=-9\mpid$
and the $q^2$ dependence is expected to be smooth in the $q^2 > 0$
region. Models of the nucleon-nucleon interaction suggest a simple
approximation for the behaviour in this region~\cite{Machleidt:1987hj},
\be\lbl{piNNmom}
g_{\pi pp}(-q^2) \simeq 
\frac{\Lambda_\pi^2}{\Lambda_\pi^2+ q^2}\, g_{\pi pp}(0)
\en
with $\Lambda_\pi\simeq 1.3$ GeV.
We can now write the $\mu p$ potential, taking into account a more complete
picture of the momentum dependence, as
\be
V_{\mu p}(q^2)=\lambda\, \frac{\Lambda_\pi^2}{\Lambda_\pi^2+ q^2}\,
\frac{{\cal A}_\mu(-q^2) \,
\vec{\sigma}_\mu\cdot \vec{q}\,
\vec{\sigma}_p \cdot \vec{q}} {q^2+\mpid}
\en  
(where $\lambda$ is given in eq.~\rf{yukawamom}.)
From this, it is not difficult to compute the Fourier transform, using the
representation~\rf{disprelA} for ${\cal A}_\mu(-q^2)$, and then the expectation
values using the formulae of the preceding section. The result for the $2S$
states can be written in the form,
\be
\braque{2S\vert V_{\mu p} \vert 2S}= \braque{\sigma_\mu\cdot\sigma_p}\,\left(
-{\cal A}_\mu(0) + \delta{\cal A}_1 + \delta{\cal A}_2
\right) \lambda\, \frac{\mpizd}{12\pi} Y_S(\mpiz)
\en
where the two corrective terms $\delta{\cal A}_1$, $\delta{\cal A}_2$ have the
following expressions
\be
\delta{\cal A}_1= {\cal A}_\mu(0)\, \frac{\mpiz}{\Lambda_\pi+\mpiz}
\en
and
\begin{align}\lbl{deltaA2}
\delta{\cal A}_2= \frac{2}{\pi}\int_0^\infty dx\, &  
\frac{\im{\cal A}_\mu(\mpizd x^2)}{x (x^2-1)}\bigg [
\frac{x^4}{1-R_\pi\,x^2} \frac{Y_S(\mpiz x )}{ Y_S(\mpiz  )}\nonumber\\ 
\ & -\frac{1}{1-R_\pi} \left(
\frac{ x^2-1}{(1-R_\pi\,x^2)R_\pi} \frac{Y_S(\Lambda_\pi)}{Y_S(\mpiz)} 
+1 \right)
\bigg]
\end{align}
with $R_\pi= \mpizd/\Lambda_\pi^2$. This expression  agrees with the result of
ref.~\cite{Hagelstein:2015lph} in the limit $\Lambda_\pi\to \infty$ and using
the leading order approximation in $\alpha$ of the function $Y_S$ (which is
valid except when $x$ is very close to zero). Fig.~\fig{FV2} shows that
the integrand in eq.~\rf{deltaA2} is peaked at $x=0$.  The effect of
$\Lambda_\pi$ is essentially to cutoff the integration region $x >
\Lambda_\pi/m_\pi$ which reduces the size of $\delta{\cal A}_2$ by 30\%
approximately. Using the numerical result~\rf{lambdatilde} for ${\cal
  A}_\mu(0)$ we find, for the two corrective terms induced by the $q^2$ dependence
of the vertices,
\be
\delta{\cal A}_1\simeq -0.52 ,\quad
\delta{\cal A}_2\simeq -2.30
\en
which reduce the result based on ${\cal A}_\mu(0)$ by roughly 50\%. It seems
reasonable to affect an uncertainty of $\simeq 30\%$ to these corrective
terms. We thus arrive at the following final estimate for the $2S$ hyperfine
splitting induced by the exchange of one pion in muonic hydrogen,
\be\lbl{2SHFSfinal}
\Delta{E}^\pi_{HFS}= -(0.09 \pm 0.06)\ \mu\rm{eV}\ .
\en
The magnitude of this result is compatible with that  obtained in
ref.~\cite{Hagelstein:2015lph} within the errors. 

\section{Conclusions}
The recent measurement of the 2S HFS in muonic
hydrogen~\cite{Antognini:1900ns} incites one to try to improve the theoretical
evaluations of the strong interaction effects, in order to reduce the error in
the determination of the Zemach radius $r_Z$. In this context, we have
considered here the ``simple'' one-pion exchange (Yukawa) contribution.  We
have indicated how to compute this contribution based on experimental results
on $\pi^0\to e^+ e^-$, $\eta\to \mu^+\mu^-$ and the associated low energy
chiral expansion as developed, in this sector, in
ref.~\cite{Savage:1992ac}. The use of chiral symmetry is important in order to
properly fix the signs of the relevant $\pi\ell\ell$ and $\pi NN$ coupling
constants and is also necessary in order to perform low-momentum expansions at
the vertices.  The final result for the contribution of one-pion exchange to
the HFS is given in eq.~\rf{2SHFSfinal}. It has a magnitude comparable to the
smallest contributions which are already taken into account in the theoretical
evaluation of the HFS (see the list of 28 contributions collected in table 3
of ref.~\cite{Antognini:2013jkc}).  At present, however, the main source of
uncertainty affecting the strong interaction effects in the 2S HFS is that
attached to the proton forward polarizabilities.

\section*{Acknowledgements:} 
We thank Vladimir Pascalutsa, 
Franziska Hagelstein and Hai-Qing Zhou for clarifying correspondence.

\bibliography{essai,atom}
\bibliographystyle{epj}

\end{document}